\documentclass[prl,twocolumn,showpacs,superscriptaddress]{revtex4-1}
\usepackage{amsmath,amssymb}
\usepackage{array}
\usepackage[T1]{fontenc}
\usepackage{bm}
\usepackage{times}
\usepackage{courier}
\usepackage{color}
\usepackage{fancyvrb}
\usepackage{graphicx}
\usepackage[colorlinks=true]{hyperref}

\newcommand\ket[1]{\ensuremath{|#1\rangle}}
\newcommand\bra[1]{\ensuremath{\langle#1|}}

\newcommand\oprod[2]{\ensuremath{|#1\rangle\langle#2|}}

\fvset{gobble=6,frame=single,framesep=2mm,commandchars=\\\{\}}

\begin{document}

%% End-Of-Header

\title{Quantum state reduction for universal measurement based
  computation}

\author{Xie Chen}%
\affiliation{Department of Physics, Massachusetts Institute of
  Technology, Cambridge, Massachusetts, USA}%
\author{Runyao Duan}%
\affiliation{Centre for Quantum Computation and Intelligent Systems,
  Faculty of Engineering and Information Technology, University of
  Technology, Sydney, NSW, Australia}%
\affiliation{Department of Computer Science and Technology, Tsinghua
  National Laboratory for Information Science and Technology, Tsinghua
  University, Beijing, China}%
\author{Zhengfeng Ji}%
\affiliation{Perimeter Institute for Theoretical Physics, Waterloo,
  Ontario, Canada}%
\affiliation{State Key Laboratory of Computer Science, Institute of
  Software, Chinese Academy of Sciences, Beijing, China}%
\author{Bei Zeng}%
\affiliation{Institute for Quantum Computing, University of Waterloo,
  Waterloo, Ontario, Canada}%
\affiliation{Department of Combinatorics and Optimization, University
  of Waterloo, Waterloo, Ontario, Canada}%

\date{Feb 8, 2010}

\begin{abstract}
  Measurement based quantum computation (MBQC), which requires only
  single particle measurements on a universal resource state to
  achieve the full power of quantum computing, has been recognized as
  one of the most promising models for the physical realization of
  quantum computers. Despite considerable progress in the last decade,
  it remains a great challenge to search for new universal resource
  states with naturally occurring Hamiltonians, and to better
  understand the entanglement structure of these kinds of states. Here
  we show that most of the resource states currently known can be
  reduced to the cluster state, the first known universal resource
  state, via adaptive local measurements at a constant cost. This new
  quantum state reduction scheme provides simpler proofs of
  universality of resource states and opens up plenty of space to the
  search of new resource states, including an example based on the
  one-parameter deformation of the AKLT state studied in [Commun.
  Math. Phys. 144, 443 (1992)] by M. Fannes {\it et al.} about twenty
  years ago.
\end{abstract}


\maketitle

Measurement based quantum computation (MBQC)~\cite{RB01}, an
interesting computation model that incorporates peculiar aspects in
quantum mechanics like entanglement and measurement, achieves the full
power of quantum computing by adaptive local measurements on a
resource state. The first MBQC scheme, also known as the one-way
quantum computation, employs the now well-known cluster
state~\cite{BR01}. The highly entangled feature of the cluster state
indicates that high entanglement is a key requirement for universality
in MBQC. Although this is true in some sense~\cite{SDV06,Vid03}, it is
also clear now that too much entanglement could also undermine
universality~\cite{GFE09,BMW09}. In other words, the entanglement
should be manageable in a structured way.

As shown in the recent breakthrough made by D. Gross {\it et al.\/} in
Refs.~\cite{GE07,GESP07}, the matrix product state (MPS)
formalism~\cite{PVMC07,FNW92} or, in higher spacial dimensions, the
computational tensor networks (CTN)~\cite{GESP07,VC04,VWPC06} provides
such an infrastructure for manipulating the entanglement and brings a
new scheme of MBQC called the correlation space quantum computation.
In this framework, a lot of new resource states beyond the cluster
state are proposed. Most of the new resource states have different
properties from the cluster state concerning, for example, local
entropy, the two-point correlation function and the locality of the
Hamiltonians of which they are unique ground states. Especially, some
of the new resources are unique ground states of more practical
Hamiltonians~\cite{BM08,CZG+09}, thereby overcoming the major flaw of
the cluster state of not being a unique ground state of any two-body
nearest-neighbour gapped Hamiltonian~\cite{Nie06}.

Here, we introduce the concept of quantum state reduction for MBQC and
the motivation is twofold. First of all, state reduction serves as a
tool for revealing the entanglement structure of universal resource
states~\cite{NDMB07,BBD+09}. Similar to the common technique to study
entanglement by considering local
transformations~\cite{BBPS96,Nie99,DVC00}, quantum state reduction is
also a type of local transformation tailored to meet the nature of
MBQC. We find out that almost all known resource states can be locally
transformed to a cluster state via quantum state reduction, indicating
that these resource states possess similar entanglement structure as
the cluster state. Compared to the attempt made in Ref.~\cite{CDV+09},
where it was shown that almost all these resource states are
``universal state preparators'', quantum state reduction is a more
direct approach and most notably more respectful for the geometry of
the resource.

\begin{figure}[htbp]
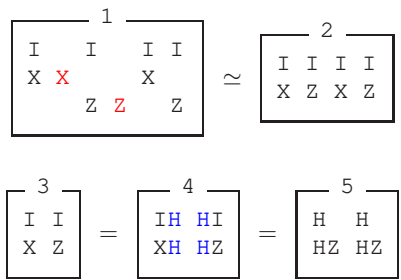

  \centering
  \begin{tabular}[c]{m{8.1em} c m{5.6em}}
    \begin{Verbatim}[label=1]
      I   I   I I
      X \textcolor{red}{X}     X  
          Z \textcolor{red}{Z}   Z
    \end{Verbatim}
    & \parbox{1.5em}{\vspace{-.8em}$\simeq$} &
    \begin{Verbatim}[label=2]
      I I I I
      X Z X Z
    \end{Verbatim}
  \end{tabular}\\
  \begin{tabular}[c]{m{3.1em} c m{4.3em} c m{4.3em}}
    \begin{Verbatim}[label=3]
      I I
      X Z
    \end{Verbatim}
    & \parbox{1.5em}{\vspace{-.8em}$=$} &
    \begin{Verbatim}[label=4]
      I\textcolor{blue}{H} \textcolor{blue}{H}I
      X\textcolor{blue}{H} \textcolor{blue}{H}Z
    \end{Verbatim}
    & \parbox{1.5em}{\vspace{-.8em}$=$} &
    \begin{Verbatim}[label=5]
      H  H
      HZ HZ
    \end{Verbatim}
  \end{tabular}\vspace{-1.5em}
  \caption{AKLT reduced to $1$-D cluster}
  \label{fig:AKLT2Cluster}
\end{figure}

Secondly, although the application of the MPS/CTN formalism in the
theory of MBQC is elegant and fruitful, the routine for analyzing the
universality of a resource state remains a complicated procedure,
including the initialization, embedding of universal rotations, the
readout, and compensation for the randomness. The quantum state
reduction approach largely simplifies the analysis. For example, the
universality of AKLT state~\cite{AKLT87,AKLT88} is now cleanly
summarized in Fig.~\ref{fig:AKLT2Cluster}. The simplicity also enables
us to find new resource states, giving the universality of two
deformations of AKLT state almost for free.

To be more precise, our state reduction is a transformation from one
resource state to some other universal target state (usually the
cluster state) using local measurement and adaptive classical control.
This transformation is named reduction as it resembles the reduction
in complexity theory---as long as $\ket{\Psi}$ is reducible to
$\ket{\Phi}$, it is in principle no harder to construct MBQC schemes
for $\ket{\Psi}$ than for $\ket{\Phi}$. It is important to note that,
although the reduction is a random procedure, it always succeeds in
obtaining a target state. Possibly, the resulting state may consist
much smaller number of particles than the state before reduction.
However, the cost or efficiency of the reduction, measured by the
diminution in the number of particles, is always expected to be a
constant.

{\em Matrix product states.---\/}Following the notion in
Refs.~\cite{GE07,GESP07}, a matrix product state $\ket{\Psi_n}$ of $n$
particles has the following form
\begin{equation}
  \label{eq:mps}
  \ket{\Psi_n} = \sum_{x_1,\cdots,x_n=0}^{d-1} \bra{R}A[x_n]\cdots
  A[x_1]\ket{L} \ket{x_1\cdots x_n}.
\end{equation}
The physical dimension of each site is $d$, while the bond dimension
of the state defined by the size of the matrices is $\delta$. In
general, one may consider an MPS where the defining matrices are
site-dependent. The defining matrices are usually far more important
than the boundary conditions $\bra{R}$, $\ket{L}$ and sometimes we
will use only the $d$-tuple $(A[0],\ldots,A[d-1])$ to specify an MPS.
We will ignore the effect of a local change of basis on the physical
space. For example, a state may be said to have some MPS
representation even though this is only true up to some local unitary
operations. Also for simplicity, $A[x]$'s will be given up to some
normalization constant, which can be figured out from
$\sum_xA[x]^\dagger A[x] = I$ shown in Ref.~\cite{PVMC07}.

A lot of states of particular interest in quantum information are
indeed matrix product states of small bond dimension. The cluster
state of one spacial dimension (i.e. a chain, or a $1$-D cluster
state), for example, is an MPS with defining matrices $(H,HZ)$, where
$H,X,Y,Z$ are used to denote the Hadamard and Pauli matrices
respectively. The GHZ state can be represented by $(I,Z)$ up to local
rotations. Another intriguing example is the AKLT
state~\cite{AKLT87,AKLT88} first studied in condensed matter theory.
It will be shown in the next section that it has a simple MPS
representation $(I,X,Z)$.

The correlation space quantum computation employs the structure of MPS
as in Eq.~\eqref{eq:mps}. It starts with the initial state $\ket{L}$
of the so called correlation space, measures the physical spaces
sequentially and thereby processes the correlation space, and finally
reads out the information stored in the correlation space. Several new
universal resources for MBQC were introduced in
Refs.~\cite{GE07,GESP07} including a modified AKLT state with defining
matrices $(H,X,Y)$. Later, the original AKLT state is also shown to be
universal for MBQC~\cite{BM08}. Recently, the concept of quantum wires
is defined and fully characterized in Ref.~\cite{GE08}, which
essentially gives the explicit condition for an MPS with $d=\delta=2$
to be a universal resource. There are two normal forms of the matrices
for quantum wires. One is the ``byproduct normal form''
\begin{equation}
  \label{eq:byproduct_form}
  A[0] = W/\sqrt{2},\; A[1] = WS(\phi)/\sqrt{2},
\end{equation}
and the other is the ``biased normal form''~\cite{Gro08}
\begin{equation}
  \label{eq:biased_form}
  A[0] = \sin\gamma\, W',\; A[1] = \cos\gamma\, W'Z,
\end{equation}
where $W,W'$ are rotations along axes in the $X$-$Z$ plane of the
Bloch sphere and $S(\phi)=\exp(-i\phi Z/2)$.

There is a higher spacial dimensional generalization of MPS, known by
the names of the computational tensor networks~\cite{GESP07}, or the
projected entangled pair state (PEPS)~\cite{VC04,VWPC06}. Due to
limitations of space, we refer the readers to the references above for
details.

{\em The tabular form.---\/}For the convenience of later discussions,
we introduce a tabular form of MPS. In the tabular form, one writes
the defining matrices of a block of sites explicitly in a table, where
each column consists the $d$ matrices of a corresponding site. The
physical indexes determine a selection of one matrix from each column,
whose product gives the correct amplitude together with the boundary
conditions $\bra{R}$ and $\ket{L}$. From the definition in
Eq.~\eqref{eq:mps} and the properties of MPS~\cite{PVMC07}, we have 1)
For any two neighboring columns, multiplication of $M$ to the right of
all matrices in the left column and $M^{-1}$ to the left of all
matrices in the right column simultaneously does not change the state;
2) A unitary transformation in the physical space corresponds to
linear combinations of entries in the column with coefficients of the
unitary; 3) Measurement in the computational basis corresponds to the
deletion of column entries not consistent with the measured outcome;
and 4) Columns of single entry can be removed by absorbing them to a
neighboring column. We will use binary relations $=,\simeq$ and
$\prec$ to represent equality, local unitary equivalence and quantum
state reduction respectively.

As an example, Table~\texttt{1} of Fig.~\ref{fig:IXZ} consists of a
block of two sites of the AKLT states which will be discussed in the
next section and it equals Table~\texttt{2} by Property 1) of the
tabular form.

{\em Reduction of the AKLT state.---\/}The AKLT
state~\cite{AKLT87,AKLT88}, named after Affleck, Kennedy, Lieb, and
Tasaki, has become one of the prototypical states of spin systems. It
also gives an excellent example for quantum state reduction.

As the origin of matrix product states, the AKLT state bears a simple
MPS representation with
\begin{equation}
  \label{eq:aklt}
  A[0] = Z, A[1] = \sqrt{2}\, \oprod{0}{1}, A[2] = \sqrt{2}\, \oprod{1}{0}.
\end{equation}
Up to a local unitary operation, the matrices of the AKLT state can
also be chosen as $(X,Y,Z)$. In fact, any three different matrices of
the identity and the Pauli matrices will work. For example,
Fig.~\ref{fig:IXZ} presents the proof that $(I,X,Z)$ also stands for
the AKLT state. In this figure, Table~\texttt{2} is obtained by adding
the $Y$'s with blue color, and hence represents the same state as
Table~\texttt{1} by Property 1) of tabular form; Table~\texttt{2} and
\texttt{3} describe two states that are equivalent under local unitary
operations by Property 2).

\begin{figure}[htbp]
  \centering
  \begin{tabular}[c]{m{3.1em} c m{4.3em} c m{3.1em}}
    \begin{Verbatim}[label=1]
      X X
      Y Y
      Z Z
    \end{Verbatim}
    & \parbox{1.5em}{\vspace{-.8em}$=$} &
    \begin{Verbatim}[label=2]
      X\textcolor{blue}{Y} \textcolor{blue}{Y}X
      Y\textcolor{blue}{Y} \textcolor{blue}{Y}Y
      Z\textcolor{blue}{Y} \textcolor{blue}{Y}Z
    \end{Verbatim}
    & \parbox{1.5em}{\vspace{-.8em}$\simeq$} &
    \begin{Verbatim}[label=3]
      Z Z
      I I
      X X
    \end{Verbatim}
  \end{tabular}\vspace{-1.5em}
  \caption{AKLT as $(I,X,Z)$}
  \label{fig:IXZ}
\end{figure}

We now show the reduction from the AKLT state to the $1$-D cluster
state. It is convenient to start with the $(I,X,Z)$ form. Two
different measurements $\mathcal{N}_1$ and $\mathcal{N}_2$ will be
used alternatively, where $\mathcal{N}_1$ measures $\{\ket{0},
\ket{1}\}$ versus $\ket{2}$, and $\mathcal{N}_2$ measures $\{\ket{0},
\ket{2}\}$ versus $\ket{1}$. That is, each measurement consists a two
dimensional and a one dimensional projectors. The measurements are
called success (failure) if the outcome corresponds to the two (one)
dimensional subspaces. We measure the two measurements sequentially
along the AKLT chain and switch the measurement we use only when the
previous one succeeds. This simple procedure is called the alternating
measurement scheme.

Table~\texttt{1} in Fig.~\ref{fig:AKLT2Cluster} denotes a possible
result after the alternating measurements on the AKLT state. More
specifically, one first measures $\mathcal{N}_1$ and succeeds. Next,
the measurement $\mathcal{N}_2$ is used. It results in the single
dimensional space $\ket{1}$ once, and succeeds subsequently, and so
on. After renaming the physical indexes and absorbing the $X$ and $Z$
in red color to their previous columns, we have Table~\texttt{2}
in Fig.~\ref{fig:AKLT2Cluster} by Property 4) and 2). This is actually
already a $1$-D cluster state by the second line of reasoning in
Fig.~\ref{fig:AKLT2Cluster}.

{\em Families of universal states of AKLT type.---\/} The simplicity
of the above analysis enables us to generalize the same approach to a
larger family of AKLT type of states. Notice that the key property
that validates the first line of Fig.~\ref{fig:AKLT2Cluster} is simply
$X^2=Z^2=I$ and the key to the second line is that $H^2=I$ and
$XH=HZ$. We now choose two unitary matrices $A$ and $B$ such that
$A^2=B^2=I$, where $A,B$ correspond to $\pi$-rotations along
$\bm{n}_a$ and $\bm{n}_b$ on the Bloch sphere respectively. Let
$C\propto A+B$ be the $\pi$-rotation along $\bm{n}_a + \bm{n}_b$. We
will have $C^2=I$ and $AC=CB$. Therefore, we can prove the following
reduction similarly
\begin{equation}
  (I,A,B)\prec(C,CB).
\end{equation}
Employing the gauge freedom of the representation of MPS, one can
always choose $B$ to be $Z$ and $C$ to be $\sin\theta X + \cos\theta
Z$, making $(C,CB)$ a quantum wire in the normal form of
Eq.~\eqref{eq:byproduct_form}. Note that the $1$-D AKLT state is a
special case where $\theta=\pi/4$ and that the error group $\langle
C,B\rangle$ is isomorphic to the dihedral group for infinitely many
$\theta$'s. With techniques of Ref.~\cite{PVMC07}, one can check that
the new AKLT type resource is always unique ground state of a
nearest-neighbor, frustration-free Hamiltonian.

In Ref.~\cite{FNW92}, Fannes, Nachtergaele and Werner considered
another one-parameter deformation of the AKLT model whose ground state
is an MPS with
\begin{equation*}
  A[0] = \sin\theta Z, A[1]= \cos\theta \oprod{0}{1}, A[2] =
  \cos\theta \oprod{1}{0}.
\end{equation*}
We will show the universality of these states also by reduction.
First, the defining matrices can be chosen as
\begin{equation*}
  \bigl(
  \sin\theta Z, \cos\theta X/\sqrt{2}, \cos\theta Y/\sqrt{2}
  \bigr),
\end{equation*}
up to local unitary transformation. Let $\hat\theta$ be the angle that
satisfies $\tan \hat\theta = \sqrt{2} \tan\theta$. The defining
matrices can be simplified to $( \sin\hat\theta I, \cos\hat\theta X,
\cos\hat\theta Z ),$ in the same way as in Fig.~\ref{fig:IXZ}. Using a
similar alternating measurement scheme in Fig.~\ref{fig:AKLT2Cluster},
this is further reducible to $( \sin\hat\theta H, \cos\hat\theta HZ
)$, a universal state in the biased normal form of
Eq.~\eqref{eq:biased_form}. One caveat is that, in this case, we
cannot simply absorb the $X$ and $Z$ in red color in
Fig.~\ref{fig:AKLT2Cluster} to neighboring sites because of the bias.
But one can always measure the computational basis in several
neighboring sites and cancel their effects by a random walk on the
Pauli group.

{\em Reduction of quantum wires to cluster states.---\/}We now discuss
the reduction of universal quantum wires to the cluster state. The
special case of $(W,WZ)$ is much easier to deal with. To transform it
into $(H,HZ)$, one can simply implement $HW^\dagger$ in the
correlation space using the sites beforehand. In the general case of
$(W,WS(\phi))$, however, one cannot succeed using projective
measurement only---the local entropy determined by $\phi$~\cite{GE08}
can never be increased. Yet, if the more general quantum measurement
is employed, this is again possible. It will be easier to work with
the biased normal form in this case. Suppose we want to transform
$(\sin\gamma W',\cos\gamma W'Z)$ to $(H,HZ)$. Assume that
$\gamma\in(0,\pi/4]$ with out loss of generality and apply on the site
a general measurement with operators
\begin{equation*}
  M_0 = \oprod{0}{0} + \tan\gamma \oprod{1}{1},
  M_1 = \sqrt{1-\tan^2\gamma}\oprod{1}{1},
\end{equation*}
known as the filtering operation. When the outcome happen to be $0$,
we have changed the matrices to $(W',W'Z)$
and can proceed as in the easy case; otherwise, we need to undo the
action of $W'Z$ on the correlation space and start all over again.
Note that it's also possible to reduce a universal quantum wire to
another quantum wire that is different from the cluster state
similarly.

{\em Higher spacial dimensional cases.---\/}This section investigates
the idea of quantum state reduction in the case of higher dimensional
resource states, which are necessary for the full power of universal
quantum computing.

The triCluster state, an interesting variant of the cluster state, is
proposed in Ref.~\cite{CZG+09} as a universal resource state of local
dimension $6$ and is the unique ground state of a two body,
frustration-free, gapped Hamiltonian. It's not difficult to see that
there is a reduction to cluster state on exactly the same lattice of
the triCluster state and we leave the details to Appendix A.

\begin{figure}[htbp]
  \centering
  \includegraphics{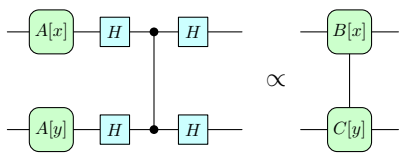}
  \caption{Reduction of 2D resources of the coupling scheme}
  \label{fig:2D}
\end{figure}

For most of the known $2$-D resource state, a general coupling scheme
has been used to make $2$-D resource states out of $1$-D chains as in
constructing the $2$-D AKLT resource, analyzing $2$-D weighted graph
state~\cite{GE07,GESP07}, and weaving quantum wires into quantum
webs~\cite{GE08}. In this scheme, one can always (a) isolate several
usually horizontal, $1$-D, universal chains from the $2$-D state and
(b) couple the correlation space of two neighboring $1$-D chains
whenever necessary. Resources of this type can be transformed to $2$-D
cluster state. To see this, one first isolate $1$-D chain states from
it; use the reductions we already have for the $1$-D case to obtain
$1$-D cluster states; and then employ an appropriate coupling to link
the $1$-D cluster states into a two dimensional cluster state. The
first two steps are obvious, while the third step is possible as shown
in Fig.~\ref{fig:2D}. In this figure, $A[x]$'s are defining matrices
of cluster state chosen to be $A[0] = \oprod{+}{0}, A[1] =
\oprod{-}{1}$, and $B$, $C$ are tensors in the notion of
Refs.~\cite{GE07,GESP07}
\begin{align*}
  B[0] & = \ket{+}_r\bra{0}_l\otimes\ket{+}_d, & B[1] & =
  \ket{-}_r\bra{1}_l\otimes\ket{-}_d,\\
  C[0] & = \ket{+}_r\bra{0}_l\otimes\ket{0}_u, & C[1] & =
  \ket{-}_r\bra{1}_l\otimes\ket{1}_u.
\end{align*}
The Hadamard gates and the CZ gate $\oprod{0}{0}\otimes I +
\oprod{1}{1}\otimes Z$, are implemented on the correlation spaces. The
right hand side of Fig.~\ref{fig:2D} represents two nodes of degree
$3$ of a cluster state.

In the weighted graph state, for example, the isolated wire has the
form $HZ^xS^z$ where $z$ depends on the outcomes of neighboring
sites~\cite{GE07,GESP07}. We can measure all sites with odd $z$ in the
$0$-$1$ basis and the resulting state is a $1$-D cluster state up to
random Clifford byproduct operations. The CZ gate can be applied in
the same way as in Ref.~\cite{GESP07}. Other coupling based schemes
mentioned above can be analyzed in a similar way.

{\em Discussions.---\/}The method of reduction for proving
universality of MBQC resource state is applicable to other examples
that have not been covered in the previous sections. These include,
for example, the modified AKLT state $(H,X,Y)$ proposed in
Ref.~\cite{GE07} and the second toric code state example in
Ref.~\cite{GESP07}. At the current stage, however, we do not know how
one can reduce the first toric code example to the $2$-D cluster
state.

It is worth comparing the idea of reduction and that of the
``universal state preparator'' result of Ref.~\cite{CDV+09}. A
reduction to cluster state would imply the universal preparator
property of the resource. On the other hand, although any universal
preparator could in principle be transformed to a cluster state, the
transformation does not respect the underlying lattice of the resource
and may be less efficient than the quantum state reduction we are
considering. For example, in the $1$-D case, the method in
Ref.~\cite{CDV+09} isolates a single particle of the physical space
while reduction of $1$-D resource results in again a $1$-D state. And
in the $2$-D case, the preparator method will need polynomial cost to
obtain a $2$-D cluster state while state reduction remains of constant
cost.

The tabular form we propose is well hinged to the structure and
properties of MPS. It simplifies the analysis by hiding the unwanted
details and provides an intuitive way of manipulating the matrices. We
have mainly investigated reductions of MPS and PEPS resources, but the
idea seems to be able to generalize to potentially new MBQC scheme not
known yet. It's also reasonable to believe that investigations of the
reduction method will improve our understanding of both the MBQC
itself and the structure of universal resource states.

\begin{acknowledgments}
RD is partly supported by QCIS, University of Technology, Sydney, and
the NSF of China (Grant Nos. 60736011 and 60702080).
ZJ acknowledges support from NSF of China (Grant Nos. 60736011 and
60721061).
BZ is supported by NSERC and QuantumWorks.
Research at Perimeter Institute is supported by the Government of
Canada through Industry Canada and by the Province of Ontario through
the Ministry of Research \& Innovation.
\end{acknowledgments}

\bibliography{mbc-reduction}

\appendix

\section{Appendix A: Reduction of triCluster state}

The triCluster state considered in Ref.~\cite{CZG+09} is a universal
resource state of local dimension $6$ and is the unique ground state
of a two body, frustration-free, gapped Hamiltonian. We give a
detailed reduction from the triCluster to cluster state in this
appendix.

\begin{figure}[htbp]
  \centering
  \includegraphics[scale=.75]{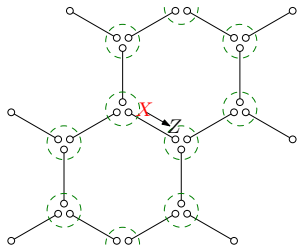}
  \caption{Reduction of triCluster state to cluster state}
  \label{fig:triCluster}
\end{figure}

The most concise way of describing the triCluster state is to employ
the PEPS picture~\cite{VWPC06}. As in Fig.~\ref{fig:triCluster}, each
bond is the state $\ket{H} \propto \ket{+}\ket{0}+\ket{-}\ket{1}$ and
each dashed circle is the projection $P = P_0 + P_1 + P_2$ where
$\ket{\pm} \propto \ket{0} \pm \ket{1}$ and
\begin{equation*}
  \begin{split}
    P_0 & = \oprod{0}{000} + \oprod{1}{111},\\
    P_1 & = \oprod{2}{001} + \oprod{3}{110},\\
    P_2 & = \oprod{4}{010} + \oprod{5}{101}.
  \end{split}
\end{equation*}

One can transform the triCluster state into a cluster state on the
same lattice, by simply measuring each site with $Q_j=P_jP_j^\dagger$
for $j=0,1,2$. Although the outcomes will be random, one can assume
that we have always measured $0$, thereby only $P_0$ are used for the
projection on each site, except that there will be some random $X$
errors happening on the bonds before the application of projections.
As the action of $X$ on one end of the bond $\ket{H}$ is equivalent to
a $Z$ on the other end, we can propagate all $X$'s to neighboring
sites as $Z$'s on the bond, which is the same as $Z$'s on the physical
space. In conclusion, we have obtained the cluster state up to some
$Z$ errors determined by the random outcomes.

\section{Appendix B: On the synchronization problem in the coupling
  method}

We have discussed the reduction for a large class of $2$-D resources
made out of $1$-D state by the coupling method. There is, however, a
tricky point that we have overlooked. This appendix aims to convey the
general idea that the problem is solvable.

\begin{figure}[htbp]
  \centering
  \includegraphics[scale=.8]{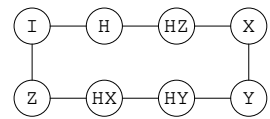}
  \caption{Random walk on the error group $\langle H,Z\rangle$}
  \label{fig:HZWalk}
\end{figure}

Imagine that we first isolate two neighboring chains, reduce them to
$1$-D cluster separately and then try to connect them into a $2$-D
cluster. Because of the randomness of the measurements, it may happen
that the two chains are not both ready for the interaction if they are
not aligned to the same column. That is, one chain contains more
unmeasured particles than the other. To solve this synchronization
problem, one may measure along these two chains in the basis that will
induce the error-group random walk. Let us consider a simple example
where the error group is $\langle H,Z\rangle$ and the interaction
implemented is the CZ gate. The random walk is the Markov chain on the
graph depicted in Fig.~\ref{fig:HZWalk}. The Markov chain is of period
$2$, meaning that it will only return to the starting state after even
number of steps. Now the measurements on the two chains induce two
independent random walk. If the difference of the number of unmeasured
particles is even, the two random walks will be in the state $I$
simultaneously after a finite number of steps. Otherwise, they will
never be $I$ simultaneously because of the periodicity of the Markov
chain. However, one can wait a finite steps for the configuration of
$I$ and $Z$. The $Z$ operation can commute with the CZ interaction and
amounts to a byproduct for future steps of the reduction.

A similar argument will work for the case where $Z$ is a generator of
the error group and the interaction implemented is CZ. As $Z$ is a
generator, the period of the random walk is either $1$ or $2$, and the
synchronization problem can be dealt with similarly.

%% Start-Of-Trailer

\end{document}